\documentclass[
reprint,
aps,
amsmath,
amssymb,
superscriptaddress,
floatfix
]{revtex4-2}

\usepackage{times}
\usepackage{graphicx}
\usepackage{bm}
\usepackage[per-mode=symbol]{siunitx}
    \DeclareSIUnit \ev{eV}
    \DeclareSIUnit \sccm{sccm}
\usepackage{braket}
\usepackage{xspace}
\usepackage{float}
\usepackage{changepage} 
\usepackage{enumitem}
\usepackage{tabularx,colortbl} 
\usepackage{multirow}
\usepackage{afterpage}
\usepackage{color}
\usepackage{soul}
\sethlcolor{yellow}
\usepackage[flushleft]{threeparttable}
\definecolor{lightGray}{rgb}{0.863, 0.863, 0.863}
\definecolor{lightOrange}{rgb}{0.933, 0.553, 0.388}
\usepackage[utf8]{inputenc}
\usepackage[T1]{fontenc}
\usepackage{comment}
\usepackage{gensymb} 
\usepackage[colorlinks=true, allcolors=blue]{hyperref}

\newcolumntype{Y}{>{\centering\arraybackslash}X} 
\newcommand{\RN}[1]{\textup{\uppercase\expandafter{\romannumeral#1}}} 

\def\g2{g^{(2)}(t)}
\newcommand{\autocorr}[1]{g^{(2)}(#1)}

\begin{document}


\title{Room Temperature Dynamics of an Optically Addressable Single Spin \\ in Hexagonal Boron Nitride}


\author{Raj N. Patel}
\thanks{These authors contributed equally}
\affiliation{
Quantum Engineering Laboratory, Department of Electrical and Systems Engineering, University of Pennsylvania, Philadelphia, PA 19104, USA
}

\author{Rebecca E. K. Fishman}
\thanks{These authors contributed equally}
\affiliation{ 
Quantum Engineering Laboratory, Department of Electrical and Systems Engineering, University of Pennsylvania, Philadelphia, PA 19104, USA
}
\affiliation{
Department of Physics and Astronomy, University of Pennsylvania, Philadelphia, PA 19104, USA
}

\author{Tzu-Yung Huang}
\altaffiliation[Present address:]{
Nokia Bell Labs, 600 Mountain Ave.,
Murray Hill, NJ 07974, USA
}
\affiliation{ 
Quantum Engineering Laboratory, Department of Electrical and Systems Engineering, University of Pennsylvania, Philadelphia, PA 19104, USA
}

\author{Jordan A. Gusdorff}
\affiliation{
Quantum Engineering Laboratory, Department of Electrical and Systems Engineering, University of Pennsylvania, Philadelphia, PA 19104, USA
}
\affiliation{
Department of Materials Science and Engineering, University of Pennsylvania, Philadelphia, PA 19104, USA
}

\author{David A. Fehr}
\affiliation{
Department of Physics and Astronomy, University of Iowa, Iowa City, IA 52242, USA
}

\author{David A. Hopper}
\altaffiliation[Present address:]{
imec, 220 Montgomery Street, Suite 1027, San Francisco, CA 94104, USA}
\affiliation{ 
Quantum Engineering Laboratory, Department of Electrical and Systems Engineering, University of Pennsylvania, Philadelphia, PA 19104, USA
}
\affiliation{
Department of Physics and Astronomy, University of Pennsylvania, Philadelphia, PA 19104, USA
}

\author{S. Alex Breitweiser}
\affiliation{ 
Quantum Engineering Laboratory, Department of Electrical and Systems Engineering, University of Pennsylvania, Philadelphia, PA 19104, USA
}
\affiliation{
Department of Physics and Astronomy, University of Pennsylvania, Philadelphia, PA 19104, USA
}

\author{Benjamin Porat}
\altaffiliation[Present address:]{
Raytheon, 2000 E El Segundo Blvd, El Segundo, CA 90245, USA
}
\affiliation{
Quantum Engineering Laboratory, Department of Electrical and Systems Engineering, University of Pennsylvania, Philadelphia, PA 19104, USA
}

\author{Michael E. Flatt\'e}
\affiliation{
Department of Physics and Astronomy, University of Iowa, Iowa City, IA 52242, USA
}
\affiliation{
Department of Applied Physics, Eindhoven University of Technology, P.O. Box 513, 5600 MB Eindhoven, The Netherlands
}

\author{Lee C. Bassett}
\email[Corresponding author. Email: ]{lbassett@seas.upenn.edu}
\affiliation{ 
Quantum Engineering Laboratory, Department of Electrical and Systems Engineering, University of Pennsylvania, Philadelphia, PA 19104, USA
}


\begin{abstract}

    Hexagonal boron nitride (h-BN) hosts pure single-photon emitters that have shown evidence of optically detected electronic spin dynamics.
    However, the electrical and chemical structure of these optically addressable spins is unknown, and the nature of their spin-optical interactions remains mysterious.
    Here, we use time-domain optical and microwave experiments to characterize a single emitter in h-BN exhibiting room temperature optically detected magnetic resonance.
    Using dynamical simulations, we constrain and quantify transition rates in the model, and we design optical control protocols that optimize the signal-to-noise ratio for spin readout. 
    This constitutes a necessary step towards quantum control of spin states in h-BN.
\end{abstract}
\maketitle



Optically interfaced solid-state spins 
enable quantum technologies with unprecedented capabilities for sensing \cite{Aslam2017,Degen2017a,Block2021,Abobeih2019Atomic-scaleSensor}, communication \cite{Hermans2022}, quantum-coherent memories \cite{Bradley2019AMinute, doi:10.1126/science.add9771}, and exploration of fundamental physics \cite{Choi2017}.
Several host materials are available \cite{Wolfowicz2021} and new ones continue to be explored in search of desirable properties \cite{Bassett2019}.
Hexagonal boron nitride (h-BN), a wide-bandgap semiconductor that hosts numerous species of optical defects, is especially promising for its low-dimensional morphology that facilitates efficient photon collection and device engineering advantages compared to three-dimensional crystals \cite{Aharonovich2022}.

A series of recent observations have confirmed the potential of h-BN as a host for quantum defects.
Room-temperature optical emitters in h-BN have shown single-photon emission \cite{Patel2022}.
Select emitters further exhibit magnetic-field-dependent photoluminescence, optically detected magnetic resonance (ODMR), and quantum-coherent spin oscillations \cite{Guo2023, Exarhos2019,Gottscholl2020,Chejanovsky2021,Stern2022}, all of which are prerequisites to establishing optically addressable spin qubits in h-BN.

Despite this progress, paramagnetic single-photon emitters are a minority of those reported on in h-BN, with recent observations noting a yield of $\sim$5\% \cite{Stern2022}. 
Emitters in h-BN exhibit heterogeneous optical and spin properties that vary dramatically even within the same sample \cite{Exarhos2017}.
Many questions therefore remain about the nature of these emitters.
Ultimately, the informed design of spin control protocols that are optimized for applications requires a detailed understanding of their optical and spin dynamics.

In this letter, we investigate an emitter in h-BN that exhibits single-photon emission and ODMR at room temperature. 
We probe the emitter's optical and spin dynamics using photon emission correlation spectroscopy (PECS) \cite{Fishman2023} and time-domain optical and microwave control.
Guided by these experiments, we develop a model for the emitter's energy-level structure, and we determine the rates that govern its optical and spin dynamics using quantitative simulations.
We design a readout protocol for the spin state that optimizes the signal-to-noise ratio (SNR).

The sample consists of a mechanically exfoliated h-BN flake ($\leq100$ nm) suspended on a patterned SiO$_{2}$/Si substrate \cite{Exarhos2017,Patel2022}.
In an area of $\approx 25 \times 25$ \textmu$\mathrm{m}^{2}$, one emitter exhibited magnetic-field-dependent photoluminescence (PL), amongst $\approx 20$ nonmagnetic emitters. 
We characterize the emitter's optical dynamics under ambient conditions using a custom-built confocal microscope \cite{Patel2022,Fishman2023}.
The emitter is illuminated with either of two continuous-wave (cw) lasers operating at 532 nm and 592 nm wavelengths, where excitation power and polarization are controlled.
Data recorded under 592 nm (532 nm) excitation are plotted in orange (green) in the relevant figures.

\begin{figure}[h!]
    \centering
    \includegraphics[width=3.375in]{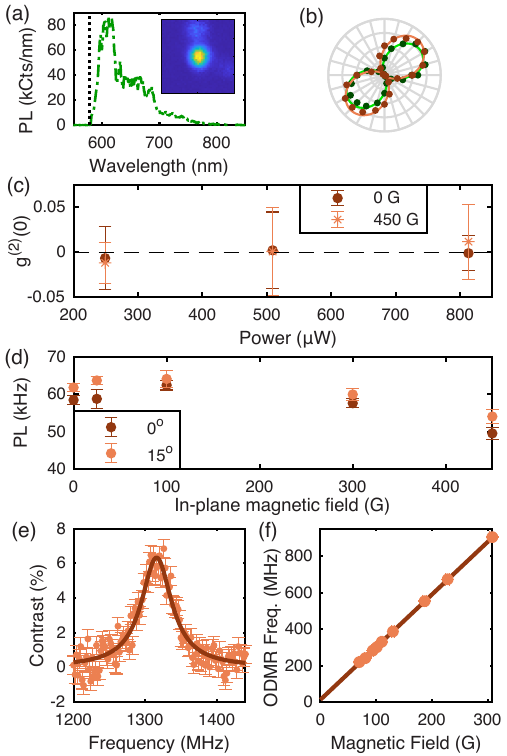}
    \caption{
    \textbf{Optical characterization and ODMR.}  
    \textbf{(a)} PL emission spectra with black dotted line representing cut-on wavelength of long-pass dichroic filter in the collection path. 
    Inset: \textmu-PL image (2$\times$2 \textmu m$^2$) of the single spin.
    \textbf{(b)} PL intensity as a function of linear excitation polarization angle for 532 nm (green circles) and 592 nm (orange circles) excitation. Solid curves are fits to the data.
    \textbf{(c)} Photon autocorrelation function at zero-delay as a function of optical power for two different in-plane magnetic fields at 0$\degree$ dipole orientation. 
    \textbf{(d)} The time-average PL emission as a function of an in-plane magnetic field for 0$\degree$ and 15$\degree$ dipole orientation. 
    \textbf{(e)} Cw ODMR spectrum (circles) at 470 G applied magnetic field and 0$\degree$ dipole orientation. 
    A Lorentzian fit (solid line) gives a resonance frequency of 1315.9 $\pm$ 0.8 MHz and a full-width half-maximum of 52 $\pm$ 2 MHz.
    \textbf{(f)} Resonance frequency measured using pulsed ODMR as a function of in-plane magnetic field.
    The solid line is a linear fit to the data.
    The x- and y-axis error bars are the same size as the data points.
    Error bars for (\textbf{e}) are propagated from Poisson error.
    All other error bars represent 68\% confidence intervals.
    }
    \label{fig:summary}
\end{figure}

Figure~\ref{fig:summary}(a) shows the emitter's PL spectrum under 532~nm excitation. 
The emitter's optical excitation is highly polarized (visibility 93 $\pm$ 3\%) and aligned for both 532 nm and 592 nm excitation (Fig.~\ref{fig:summary}(b)).
The emitted PL is polarized along the same axis (Supplemental Material, Fig.~\ref{fig:EmissionPolarization}).
While previous observations have noted  heterogeneous polarization responses for h-BN's emitters \cite{Jungwirth2016,Exarhos2017,Jungwirth2017,Ziegler2018,Patel2022}, indicating the presence of multiple electronic excited states, the aligned excitation and emission dipoles observed for this emitter are consistent with a single radiative excited state.
Figure~\ref{fig:summary}(c) shows the second-order photon autocorrelation function, $\autocorr{\tau}$, at zero-delay ($\tau = 0$), in the presence and absence of an applied magnetic field.
In both cases, we observe noise-limited photon antibunching, $g^{(2)}(0)=0$, independent of optical excitation power.


This emitter's PL intensity is modulated by applied dc and ac magnetic fields.
Magnetic fields affect a paramagnetic defect's PL intensity due to spin-selective transition rates that govern its optical dynamics, although this can occur in different ways \cite{Doherty2013,Exarhos2019,Fishman2023,Epstein2005}.
We apply a dc magnetic field parallel to the hBN surface and rotate the sample about the optical axis to vary the relative orientation of the optical excitation dipole to the field axis (referred hereafter as dipole orientation).
As observed in Fig.~\ref{fig:summary}(d), the steady-state PL varies by 15$\%$ on increasing the magnetic field strength from 0 G to 470 G for both 0$\degree$ and 15$\degree$ dipole orientations.
Accompanying variations in PECS measurements (Supplemental Material, Fig.~\ref{fig:opticalDynamics}) confirm that the PL changes result from magnetic modulation of the emitter's optical dynamics.
Figure~\ref{fig:summary}(e) shows an example of an ODMR spectrum acquired as a function of applied microwave frequency.
The microwaves are amplitude modulated at 12.5 kHz, and the ODMR spectrum is normalized by dividing the signal PL (microwaves on) by the reference PL (microwaves off).

Figure~\ref{fig:summary}(f) shows the best-fit ODMR center frequency as a function of applied magnetic field.
A linear fit gives $g$-factor $g = 2.06 \pm 0.06$, with a zero-field-splitting (ZFS) of 15 $\pm$ 13 MHz.
Measurements at various dipole orientations (Supplemental Material, Fig.~\ref{fig:ZFS}) indicate an isotropic $g\approx 2$, consistent with the free-electron $g$-factor, and an average ZFS of 9 $\pm$ 10 MHz, consistent with zero within experimental uncertainty.
The scale of the ZFS is consistent with prior reports, although interpretations of its significance have varied \cite{Chejanovsky2021,Stern2022}.
We observe no additional resonances at higher frequencies up to 4.2 GHz.
Based on these observations, we postulate a doublet ($S=\frac{1}{2}$) spin state.



\begin{figure}[t!]
    \centering
    \includegraphics[width=3.375in]{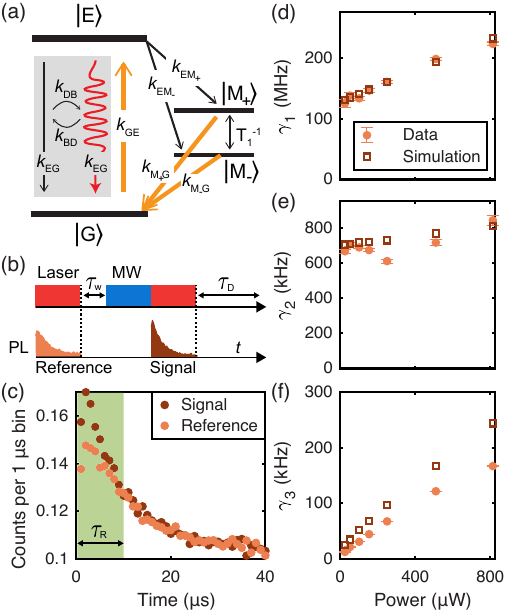}
    \caption{
    \textbf{Energy-level model and optical dynamics.}      
    \textbf{(a)} Energy-level model made up of a singlet ground and excited state, and a metastable doublet. The gray box highlights fluctuations between a nonradiative decay path (black arrow) and radiative decay (red, wavy arrow). Orange arrows depict excitation-power-dependent transitions.
    \textbf{(b)} Pulse protocol for optical spin contrast measurement. "Reference" and "Signal" denote timing windows during which photons are counted.
    \textbf{(c)} PL counts for 1 \textmu s time bins corresponding to signal (dark orange circles) and reference (light orange circles) readout, at 350 \textmu W optical power, 20 \textmu s wait time, and 40 \textmu s microwave pulse.
    Green highlights the region where contrast is observed with $\tau_\mathrm{R}$ denoting a readout window.
    \textbf{(d)}-\textbf{(f)} PECS measurements.
    \textbf{(d)} The antibunching rate, $\gamma_{1}$, 
    and the bunching rates,
    \textbf{(e)} $\gamma_{2}$ and 
    \textbf{(f)} $\gamma_{3}$, as a function of optical excitation power at 0 G magnetic field.
    }
    \label{fig:odmr}
\end{figure}

Figure~\ref{fig:odmr}(a) shows the proposed model explaining the observed optical dynamics.
The model features a metastable spin-1/2 doublet system ($|\mathrm{M}_{\pm}\rangle$) coupled to a spinless manifold of ground ($|\mathrm{G}\rangle$) and optically excited ($|\mathrm{E}\rangle$) states. 
We further identify a stochastic modulation of the optical decay pathway, fluctuating between a raditative and non-radiative transition.
Arrows denote transitions with corresponding rates, $k_{ij}$, between states $i$ and $j$, with the spin relaxation rate labelled $T_{1}^{-1}$.
The number and arrangement of levels are determined by a series of experiments and corresponding simulations.
PECS measurements (discussed later), show clear evidence of photon bunching associated with metastable dark configurations.
PECS experiments also reveal the nature of the transition mechanisms between these configurations \cite{Fishman2023}.
In addition to the optical excitation transition rate, $k_\mathrm{GE}$, PECS simulations imply that the rates  $k_\mathrm{M_{+}G}$ and $k_\mathrm{M_{-}G}$ each feature a power-dependent component in addition to a spontaneous (power-independent) component.
The rates extracted from our simulations are further supported by an analytical, three-level approximation \cite{Berthel2015}.

We propose that the spin doublet state observed in ODMR exists in a metastable configuration.
This arrangement cannot be determined by ODMR alone. 
Rather, it is confirmed by an optical spin contrast experiment, shown in Fig.~\ref{fig:odmr}(b-c), which distinguishes between configurations where the spin states exist in the optical excitation/emission manifold compared to the metastable configuration.
As shown in  Fig.~\ref{fig:odmr}(b), the laser is modulated on and off, with a dark time of duration $\tau_\mathrm{D}$, and a microwave pulse can be applied during the dark time following a wait time, $\tau_\mathrm{w}$.
Figure~\ref{fig:odmr}(c) shows the PL as a function of time during the laser pulse in situations when the microwaves are applied (signal) or not (reference).

We consider in turn the expected dynamics for configurations with spin doublets in the optical ground and excited states compared to the configuration shown, with a metastable doublet.
The former case (spins in optical manifold) would predict zero initial spin contrast when the laser is turned on, with a contrast evolving during the pulse due to spin-dependent decay rates, whereas the latter configuration (metastable spin) predicts a nonzero initial contrast that decays as the system returns to the steady state (see Supplemental Material for additional discussion). Our observations are consistent with the latter case.
Informed by classical rate equation simulations of a spin contrast experiment for the model shown in Fig.~\ref{fig:odmr}(a), we estimate a spin relaxation rate of $T_{1}^{-1} \sim$ 0.01 MHz to achieve a similar initial contrast (7$\%$) in simulations.

After establishing the main features of the electronic level structure, we next consider the rates that govern its optical dynamics.
Using PECS, we acquire $\autocorr{\tau}$ at various optical excitation powers.
From the Akaike information criterion and reduced chi-squared statistics, we determine the best-fit empirical function to be a three-timescale model,
\begin{equation}
    \autocorr{\tau} = 1 - C_{1}e^{-\gamma_{1}|\tau|} + C_{2}e^{-\gamma_{2}|\tau|} + C_{3}e^{-\gamma_{3}|\tau|},
    \label{eq:G2}
\end{equation}
where $\tau$ is the delay time, $\gamma_{1}$ and $C_{1}$ are the antibunching rate and amplitude, $\gamma_{2}$ and $\gamma_{3}$ are the bunching rates, and $C_{2}$, $C_{3}$ are the associated bunching amplitudes \cite{Patel2022}.
Both $\gamma_{1}$ and $\gamma_{3}$ increase monotonically as a function of optical excitation power (Figs.~\ref{fig:odmr}(d) and \ref{fig:odmr}(f)), as expected for processes involving optical pumping to the excited state.
In contrast, $\gamma_{2}$ shows no clear trend with respect to power (Fig.~\ref{fig:odmr}(e)).

Each PECS rate originates from transition pathways between multiple electronic states that define a distinct process.
In this model, $\gamma_{1}$ corresponds to optical excitation at rate $k_\mathrm{GE}$ followed by relaxation back to $|\mathrm{G}\rangle$ at rate $k_\mathrm{EG}$.
For a direct optical transition between two electronic states, as in this model, the antibunching rate is given by $\gamma_{1} \approx k_\mathrm{GE}+k_\mathrm{EG}$ \cite{Berthel2015, Fishman2023}.
Since $k_\mathrm{GE}$ is proportional to the optical excitation power, $p$, whereas $k_\mathrm{EG}$ is power-independent, we expect a linear variation, $\gamma_1=k_\mathrm{EG}+\beta p$, where $\beta$ is a proportionality constant.
The observations in Fig.~\ref{fig:odmr}(d) are consistent with this expectation, and a linear fit (Fig.~\ref{fig:simResults}) yields $k_\mathrm{EG}$ = 128 $\pm$ 2 MHz and $\beta = (0.125 \pm 0.005$~MHz/\textmu W$)$.

The bunching rates $\gamma_2$ and $\gamma_3$ represent processes through which the system enters nonradiative configurations.
The process of optical excitation to $|\mathrm{E}\rangle$ followed by nonradiative decay through $|\mathrm{M}_{\pm}\rangle$ back to $|\mathrm{G}\rangle$ depends on $p$ through the first step at rate $k_\mathrm{GE}$ and through the last step at rates $k_\mathrm{M\pm G}$; hence the corresponding bunching rate should increase with $p$.
Based on our observations, $\gamma_3$ is consistent with this process.
In contrast, $\gamma_2$ does not vary significantly with $p$.
To account for this unusual observation, we propose a fluctuating relaxation mechanism from $|\mathrm{E}\rangle$ to $|\mathrm{G}\rangle$ that stochastically switches between radiative and non-radiative configurations at rates $k_{\mathrm{DB}}$ and $k_{\mathrm{BD}}$ through a process independent of $p$.
This is potentially due to fluctuations in the state of a nearby coupled defect that modulates the emission process. 
As shown by the simulations in Fig.~\ref{fig:odmr}(e), this process leads to bunching in $\autocorr{\tau}$ at a nearly constant rate $\gamma_2\sim k_{\mathrm{DB}} + k_{\mathrm{BD}}$, which closely matches the experimental observations.


\begin{figure}[t!]
    \centering
    \includegraphics[width=3.375in]{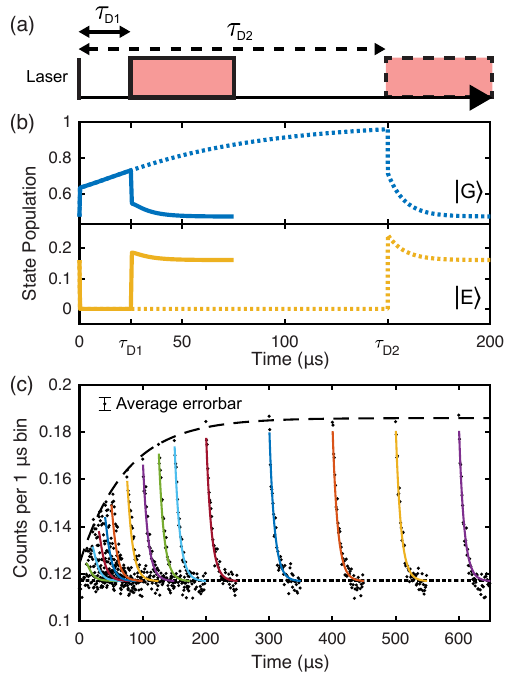}
    \caption{
    \textbf{Fluorescence recovery.}      
    \textbf{(a)} Pulse protocol for time-domain fluorescence recovery measurement.
    \textbf{(b)} Example of time-domain ground (blue lines) and excited (yellow lines) state populations during a fluorescence recovery experiment for short ($\tau_\mathrm{D1}$) and long ($\tau_{\mathrm{D2}}$) dark times between laser pulses, shown as solid and dashed lines respectively.
    \textbf{(c)} Time-domain PL as a function of dark time between optical pulses, $\tau_{D}$, acquired at 
    91 G in-plane magnetic field and 34.2$\degree$ dipole orientation, using 1 \textmu s time bins. Black dots are data and colored lines are simulation. The average uncertainty of each data point is $\pm$ 0.003 counts per bin.
    }
    \label{fig:DSR}
\end{figure}

While $\gamma_{3}$ as derived from PECS yields estimates for the overall rates connecting $|\mathrm{E}\rangle$, $|\mathrm{M}_{\pm}\rangle$, and $|\mathrm{G}\rangle$, additional measurements are required to resolve the contributions of each individual rate.
The fluorescence recovery protocol (Fig.~\ref{fig:DSR}(a)) involves varying the dark time, $\tau_\mathrm{D}$, between laser pulses and recording the time-domain PL emission during each pulse.
Figure~\ref{fig:DSR}(b) illustrates the evolution of populations in $|\mathrm{G}\rangle$ and $|\mathrm{E}\rangle$ during a fluorescence recovery experiment for two different values of $\tau_\mathrm{D}$.
The measured PL (shown in Fig.~\ref{fig:DSR}(c) for 15 different values of $\tau_\mathrm{D}$) is proportional to the population in $|\mathrm{E}\rangle$, which depends in turn on the population in $|\mathrm{G}\rangle$ at the beginning of the laser pulse.
The PL at the start of the laser pulse (dashed line in Fig.~\ref{fig:DSR}(c)) increases as a function of $\tau_\mathrm{D}$, with an extracted time constant of $73 \pm 5$ \textmu s.
This time constant directly reflects the $p$-independent components of $k_{\mathrm{M-G}}$ and $k_{\mathrm{M_{+}G}}$.
During the laser pulse, the PL decays to a steady-state value with a decay constant of $7.9 \pm 0.4$ \textmu s.
The decay constant reflects the $p$-dependent $k_{\mathrm{M-G}}$ and $k_{\mathrm{M_{+}G}}$ in addition to $k_{\mathrm{EM_{-}}}$, $k_{\mathrm{EM_{+}}}$, $k_{\mathrm{BD}}$, and $k_{\mathrm{DB}}$.
The maximum contrast from the initial to the steady-state PL, measured to be $62.2 \pm 0.2\%$, constrains the ratio $(k_{\mathrm{M_{+}G}}+k_{\mathrm{M_{-}G}})/(k_\mathrm{EM_{-}}+k_\mathrm{EM_{+}})$.

We use a rate equation model for the level configuration shown in Fig.~\ref{fig:odmr}(a) to simulate the state populations for PECS and the fluorescence recovery pulse protocol (see Supplemental Material for simulation details).
The parameters $k_\mathrm{EG}$ and $k_\mathrm{GE}$ are known based on earlier considerations, as is the sum $k_\mathrm{DB} + k_\mathrm{BD}$.
We determine the remaining parameters by empirically matching simulations of PECS rates (squares in Fig.~\ref{fig:odmr}(c-e)) and bunching coefficients (see Supplemental Material, Fig.~\ref{fig:simResults}), spin contrast (Fig.~\ref{fig:odmr}(c)), and fluorescence recovery PL (colored lines in Fig.~\ref{fig:DSR}(c)) to the data.
The simulated fluorescence recovery time constants match those extracted from the data within fit errors, and the maximum contrast matches within 5$\%$.
Slight discrepancies between the PECS simulations and data are attributed to variations of the spin-dependent rates through $|\mathrm{M}_{\pm}\rangle$ due to different applied magnetic fields for the PECS (no field) and fluorescence recovery (91 G) experiments.
The optimized rates are as follows: $k_\mathrm{EM} \equiv k_\mathrm{EM_{-}}+k_\mathrm{EM_{+}} = 0.19$ MHz, $k_\mathrm{EM_{+}}/ k_\mathrm{EM_{-}} = 0.1$, $k_\mathrm{MG} \equiv (k_{\mathrm{M_{+}G}}+k_{\mathrm{M_{-}G}})/2 = 0.034~\mathrm{MHz} + (0.8~\frac{\mathrm{MHz}}{\mathrm{mW}})p$,  $k_{\mathrm{M_{+}G}} / k_{\mathrm{M_{-}G}}$ = 8, $k_\mathrm{BD}$ = 0.15 MHz, and $k_\mathrm{DB} = 0.55$ MHz. 
More information regarding the process of quantifying transition rates can be found in the supplemental material.


\begin{figure}[b!]
    \centering
    \includegraphics[width=3.375in]{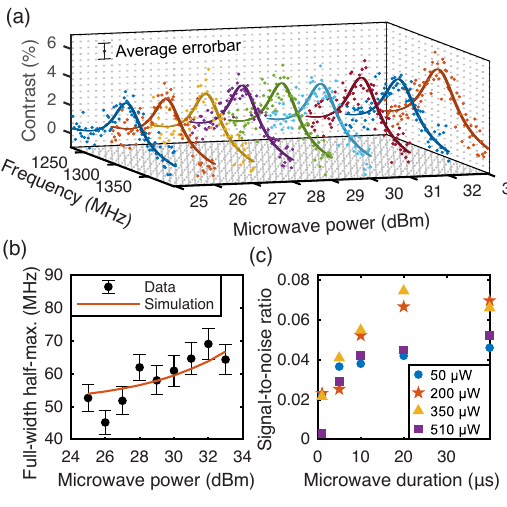}
    \caption{
    \textbf{Spin properties.}      
    \textbf{(a)} ODMR contrast data with simulation fits (lines) as a function of microwave frequency and power.
    The average uncertainty of each data point is $\pm 0.5\%$ ODMR contrast.
    \textbf{(b)} Full-width half-maximum as a function of microwave power from Lorentzian fits of simulation and data.
    \textbf{(c)} Signal-to-noise ratio as a function of microwave pulse duration for readout time, $\tau_{\mathrm{R}} = 5$ \textmu s, at varying optical powers.
    All data are acquired at 592 nm excitation.
    }
    \label{fig:Spin}
\end{figure}

In order to model the ODMR experiments, we use the Lindblad framework to capture coherent evolution of the spin states along with semiclassical optical dynamics \cite{10.1093/acprof:oso/9780199213900.001.0001}.
Our experiments indicate a variation of the ODMR contrast and linewidth as a function of microwave power, $p_\mathrm{MW}$ (Fig.~\ref{fig:Spin}(a)).
We fit Lindblad simulations to these data, using the optimized optical-dynamics rates as fixed parameters. Free parameters include the microwave coupling efficiency, $\eta$, which determines the power-dependent Rabi frequency according to $\Omega_\mathrm{R}/(2\pi)=\eta \sqrt{p_\mathrm{MW}}$, as well as the spin dephasing time, $T_{2}^{*}$.
The fits are plotted along with the data in Fig.~\ref{fig:Spin}(a), and Fig.~\ref{fig:Spin}(b) compares the ODMR linewidth extracted from the data to the best-fit simulation. 
Accounting for uncertainties in the fit and in $T_1$, we find $\eta$ = 0.0189 $\pm$ 0.0007 MHz/$\sqrt{\mathrm{W}}$ and $T_{2}^{*}$ = 6.3 $\pm$ 0.1 ns (see Supplemental Material for simulation details).


Using this quantitative understanding of the emitter's optical dynamics, we can design optimized protocols for spin initialization and readout.
Spin polarization develops under optical illumination due to the spin-dependent branching ratios $k_\mathrm{EM_{+}}/ k_\mathrm{EM_{-}}$ and $k_{\mathrm{M_{+}G}} / k_{\mathrm{M_{-}G}}$.
Our model implies a steady-state population ratio $|\mathrm{M}_{-}\rangle$:$|\mathrm{M}_{+}\rangle\sim 30$:1 at $p=350$ \textmu W.
The corresponding spin polarization $\sim97$~\% significantly exceeds the steady-state polarization for other spin defects such as the diamond nitrogen-vacancy (NV) center.
However, the population develops over $\sim10$~\textmu s, which is an order of magnitude longer than typical NV-center initialization times \cite{Hopper2018SpinDiamond}.

To optimize spin readout, we consider the single-shot SNR in a spin contrast experiment (Fig.~\ref{fig:odmr}(b)), given by~\cite{Hopper2018SpinDiamond}
\begin{equation}
    \mathrm{SNR} = \frac{\alpha_1 - \alpha_0}{\sqrt{\alpha_1+\alpha_0}}.
    \label{eq:snr}
\end{equation}
Here $\alpha_{1}$ and $\alpha_{0}$ are, respectively, the mean number of detected photons for the signal (microwave on) and reference (microwave off) recorded in a given readout window, $\tau_\mathrm{R}$ (green region in Fig.~\ref{fig:odmr}(c)).
Figure~\ref{fig:Spin}(c) shows the SNR as a function of microwave pulse duration, for various settings of $p$. 
We find an optimum SNR $\approx 0.07$ for $p=350$~\textmu W and $\tau_\mathrm{R}=5$~\textmu s (See Fig.~\ref{fig:snr} for additional measurements).
Since the spin contrast experiment in Fig.~\ref{fig:odmr}(b) compares the polarized spin configuration with a fully mixed state, the observed SNR is approximately half of what would be expected for a full spin inversion.
For comparison, the optimized spin-readout SNR for a diamond NV-center with similar photon count rate is only $\sim0.03$ \cite{Hopper2018SpinDiamond}.


The quantitative model presented in this work will directly facilitate the use of single spins in h-BN for quantum technologies. 
The inferred spin polarization of $\sim97$\% and spin-readout SNR of $\sim0.15$ are superior to the performance of well-established room-temperature spin qubits, including NV centers.
More sophisticated initialization and readout protocols could offer further improvements \cite{Hopper2018SpinDiamond}.
The spin relaxation time, $T_1\approx100$ \textmu s, is comparable to the spin lifetimes of NV centers in nanodiamonds, offering substantial opportunities for relaxometry imaging and chemical sensing.
The relatively short dephasing time, $T_{2}^{*}$ = 6.3 $\pm$ 0.1 ns, likely reflects substantial hyperfine coupling to nearby nuclear spins.
Hence, with the design of optimized microwave antennas to drive faster spin rotations, it will be possible to use dynamical decoupling protocols to substantially extend the electron-spin coherence time, and to address the states of coupled nuclear spins.
In contrast to most spin qubits, which feature spin levels in the ground and optically excited states, the spinless ground state configuration of this system can be beneficial to protect the coherence of nuclear spin states \cite{Lee2013}.

The chemical structure of h-BN's visible emitters remains a mystery.
Conclusive identification is needed to enable the further optimization of materials, devices, and quantum control protocols. 
The detailed empirical understanding of their energy-level structure and dynamics developed through this work will inform and constrain future theoretical models.
More generally, the framework followed in this letter can be used to characterize and control the optical and spin dynamics of single spins in any solid-state host material.


This work was primarily supported by the National Science Foundation (NSF) award DMR-1922278 (Penn) and DMR-1921877 (Iowa).
J.A.G. is supported by an NSF Graduate Research Fellowship (DGE-1845298).
S.A.B. acknowledges support from an IBM PhD Fellowship.
The authors gratefully acknowledge use of facilities and instrumentation in the Singh Center for Nanotechnology at the University of Pennsylvania, supported by NSF through the National Nanotechnology Coordinated Infrastructure (NNCI; Grant ECCS-1542153) and the University of Pennsylvania Materials Research Science and Engineering Center (MRSEC; DMR-1720530).
We gratefully acknowledge fruitful discussions with M. Turiansky and C. G. Van de Walle.


\bibliography{references.bib}

\clearpage




\section*{Supplemental Material}


\newcommand{\summaryFig}{1}
\newcommand{\odmrFig}{2}
\newcommand{\DSRFig}{3}
\newcommand{\spinFig}{4}
\newcommand{\autocorrEq}{1}


\section{Sample preparation} 
\label{sec:sample}

We used a custom-built confocal microscope to study individual emitters in h-BN under ambient conditions \cite{Patel2022}. 
The h-BN samples consisted of bulk, single crystals purchased from HQ Graphene.
The h-BN bulk crystals were mechanically exfoliated in thin ($\SI{\leq100}{\nano\meter}$) and large area ($\SI{\sim10}{\micro\meter}$) flakes using a dry transfer process \cite{Huang2015} and transferred to a SiO$_{2}$/Si substrate patterned with circular trenches \cite{Exarhos2017}.
Prior to the optical studies, the exfoliated h-BN samples were cleaned with a soft O$_2$ plasma (Anatech SCE 106 Barrel Asher, \SI{50}{\W} of power, \SI{50}{\sccm} O$_2$ flow rate) for 5 minutes to remove polymer residues resulting from the transfer process. 
The samples were then annealed in a tube furnace at \SI{850}{\celsius} in low flow Ar atmosphere for 2 hours.
We have found this annealing conditions to create stable emitters. 
We find the single spin to be extremely stable in ambient conditions with optical pumping up to 500 \textmu W and microwave pulses with RF power up to 4 W for over hundreds of hours.

\section{Experimental setup}

\begin{figure*}[htb!]
    \renewcommand{\thefigure}{S\arabic{figure}}
    \renewcommand\figurename{Figure }
    \let\nobreakspace\relax
    \setcounter{figure}{0}
    \centering
    \includegraphics[width=7in]{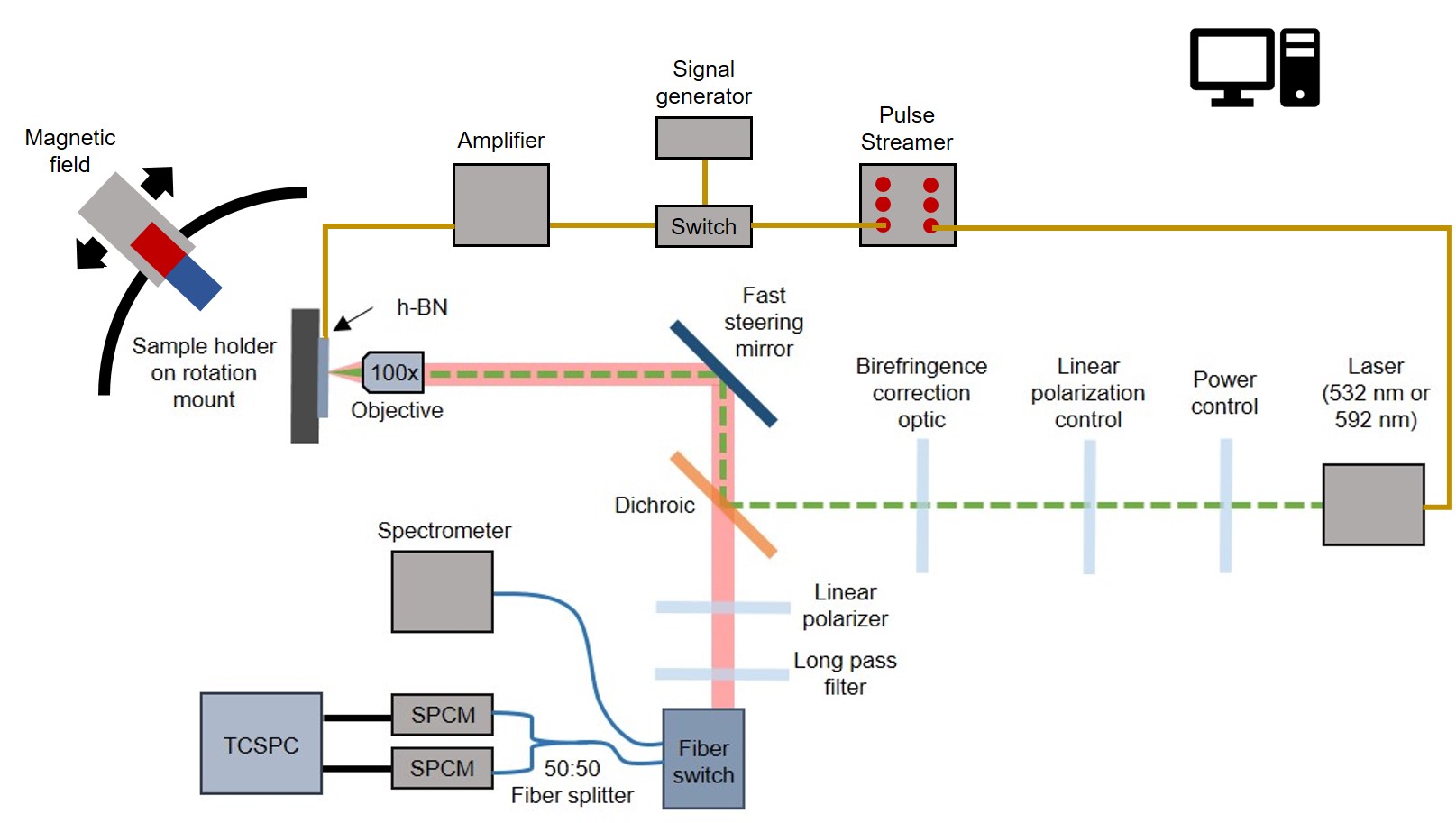}
    \caption{
    \textbf{Experimental setup.}     
    }
    \label{fig:setup}
\end{figure*}

\subsection{Optics, detection system and magnetics} 

Figure~\ref{fig:setup} depicts a simplified schematic of the room-temperature confocal microscope used to measure the emitters.
There are two available excitation sources: a 532 nm (green) cw laser (Coherent, Compass 315M-150) and a 592 nm (orange) cw laser (MPB Communications, VF-P-200-592).
The power and polarization of each excitation path can be independently selected.
Reported power values are measured just prior to the objective.
In addition, a shutter completely blanks the excitation source when imaging is not in use to mitigate unnecessary light exposure.
The excitation paths are combined with the collection path using a long pass (LP) dichroic mirror (Semrock, BrightLine FF560-FDi01 for green and Semrock, BrightLine FF640-FDi01 for orange).
The LP dichroic cut-off is 560 nm for green excitation and 640 nm for orange excitation. 
A fixed half-wave plate in each of the excitation paths corrects for the birefringence induced by the dichroic mirrors.
The co-aligned excitation and collection paths are sent through a $4f$ lens system with a fast steering mirror (Optics in Motion, OIM101) and a 0.9 NA 100x objective (Olympus, MPI Plan Fluor) at the image planes.
This allows for the collection of wide-field, rastered, micro-photoluminescence (\textmu-PL) images.
The objective is mounted on a stage system for changing the field of view.

The collection path consists of a linear polarizer (Thorlabs, WP25M-VIS) for measuring the emission polarization as well as a wide-band variable retarder (Meadowlark, LRC-100) which compensates for the birefringence induced by the dichroic.
A LP filter specific to the excitation color fully extinguishes any scattered excitation light and the Raman signal.
The cut-on wavelengths are 578 nm (Semrock, BLP01-568R-25) and 650 nm (Semrock, BLP01-635R-25) for green and orange, respectively.
The filtered light is focused onto the core of a 50 \textmu m core multi-mode fiber (Thorlabs, M42L01) acting as a pinhole.
The output of the fiber is connected to a fiber switch (DiCon, MEMS 1x2 Switch Module) which can switch the collected emission to either a 50:50 visible fiber splitter (Thorlabs, FCMM50-50A-FC) or a spectrometer (Princeton Instruments, IsoPlane160 and Pixis 100 CCD).
The outputs of the fiber splitter are sent to two identical single-photon counting modules (SPCM, Laser Components, Count T-100) resulting in a Hanbury Brown Twiss interferometer.
The outputs of the SPCMs are either measured by a data acquisition card (National Instruments, DAQ6323) for general-purpose counting or a time-correlated single-photon counting (TCSPC) module (PicoQuant, PicoHarp 300) for recording the photon time-of-arrival information with a full system resolution of $\sim$350 ps.

The magnetic field is applied using a neodymium magnet (K\&J Magnetics DY0Y0-N52) mounted on to a linear stage (Zaber Technologies T-LSR160D) that enables magnetic field strength variation. 
The stage is mounted on a home-build goniometer that allows variation in magnetic field orientation from 0$\degree$ to 90$\degree$, where 0$\degree$ corresponds to an in-plane applied magnetic field with respect to the sample plane.
The magnetic field strength as a function of stage position and orientation on the goniometer is calibrated at the sample using a hall probe (LakeShore 425 Gaussmeter).
The available magnetic field strength is 0 G to 470 G.
Magnetic field calibration was performed at each dipole orientation presented in the main text and supplemental material to account for the shift in the distance of the SPE from the magnet on rotating the sample.
This shift is up to 2 mm in X and Y direction, enough to alter the effective magnetic field by several gauss.
With the sample removed, the gaussmeter is carefully placed at the same position aided by the laser beam coming out of the objective.
The objective position corresponds to that of each orientation.
Multiple calibrations are performed to determine the systematic error in positioning the gaussmeter.
The magnetic field calibration error is taken into account in the data presented in Fig.~\ref{fig:summary}.

\subsection{Microwave instrumentation and timing electronics}

\begin{figure*}[ht!]
    \renewcommand{\thefigure}{S\arabic{figure}}
    \renewcommand\figurename{Figure }
    \let\nobreakspace\relax
    \centering
    \includegraphics[width=7in]{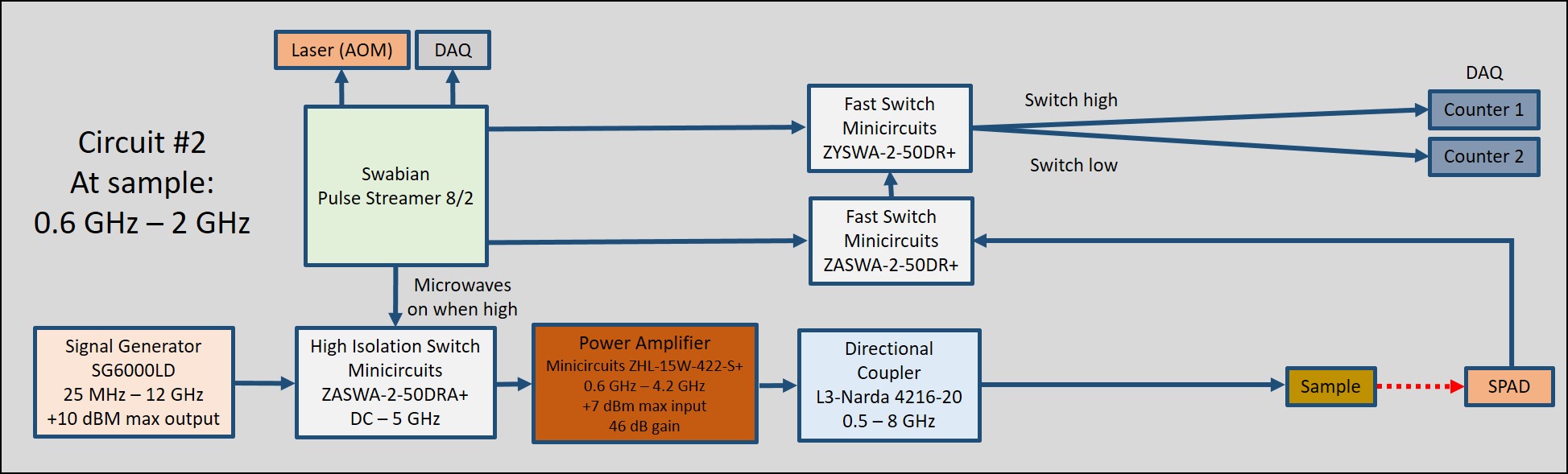}
    \caption{
    \textbf{Microwave circuit.}     
    }
    \label{fig:microwavecircuit}
\end{figure*}

Figure~\ref{fig:microwavecircuit} depicts a schematic of the RF instrumentation used for spin dynamics measurements.
An arbitrary waveform generator (Swabian Pulse Streamer 8/2) is used for optical and microwave pulse protocols by syncing timings and outputs of various electronics.
A signal generator (DS Instruments SG6000LD) is used as a source of microwaves.
The arbitrary waveform generator (AWG) is used to modulate the microwaves via a high isolation switch (Minicircuits ZASSWA-2-50DRA+).
The microwave signal is further amplified using a power amplifier (Minicircuits ZHL-20W-13+ or ZHL-15W-422-S+).
A directional coupler (L3-Narda 4216-20) at the output of the power amplifier allows for monitoring the input microwave pulses on an oscilloscope (Tektronix TDS 2024).
A custom-made microwave chip is connected to the output of the directional coupler at one end and to a 50 ohm terminator at another end. 
The patterned substrate consisting of the h-BN flake is glued to the microwave chip using rubber cement.
Using a wire bonder (Kulicke and Soffa 4523), two bonding pads on the microwave chip are connected with a thin aluminum wire that allows for the transmission of the microwaves.
The aluminum wire passes over the h-BN flake and is $\sim$50 \textmu m away from the spin-defects.
The SPCM used for the optical readout is connected to two fast switches (Minicircuits ZYSWA-2-50DR+) connected in series.
The two switches are controlled by the AWG.
The first switch referred to as the counting switch is used to send the signal coming from SPCM to next switch for recording or to a 50 ohm terminator if discarding.
The second switch referred to as the routing switch takes the signal from the counting switch and routes it to one of the two counters on the data acquisition card.
The two counters are devoted to collecting either the signal PL or the reference PL in the measurements.
The optical pulse is modulated via an acousto-optic modulator (Isomet AOM 1250C) connected to the AWG.
An output of the AWG connected to the data acquisition card is used as a clock reference.

\subsection{Microwave chip}

The final sample consisted of a substrate glued to a custom-designed microwave chip with SMA connectors as shown in Fig.~\ref{fig:microwavechip}.
Prior to that, the substrate consisting of h-BN flakes was annealed as discussed in Supplemental Sec.~\ref{sec:sample}.
The bonding pads labeled 4 and 5 on the microwave chip are connected by a thin aluminum wire using a wire bonder (Kulicke and Soffa 4523).
The wire bonding is done carefully such that the aluminum wire is as close as possible to quantum emitters (QEs) of interest that had been pre-characterized using PECS as discussed in Supplemental Sec.~\ref{sec:field-dependent-PECS}.
The chip is secured to a threaded adapter which is mounted in the setup on a rotation stage.
We control full 360$\degree$ rotation of the optical excitation dipole (referred hereafter as dipole) by controlling orientation of the sample mounted on the rotation stage.
The chip is connected to the microwave circuit shown in Fig.~\ref{fig:microwavecircuit}, with one SMA connector 50 $\Omega$ terminated.

\begin{figure}[htb!]
    \renewcommand{\thefigure}{S\arabic{figure}}
    \renewcommand\figurename{Figure }
    \let\nobreakspace\relax
    \centering
    \includegraphics[width=3.375in]{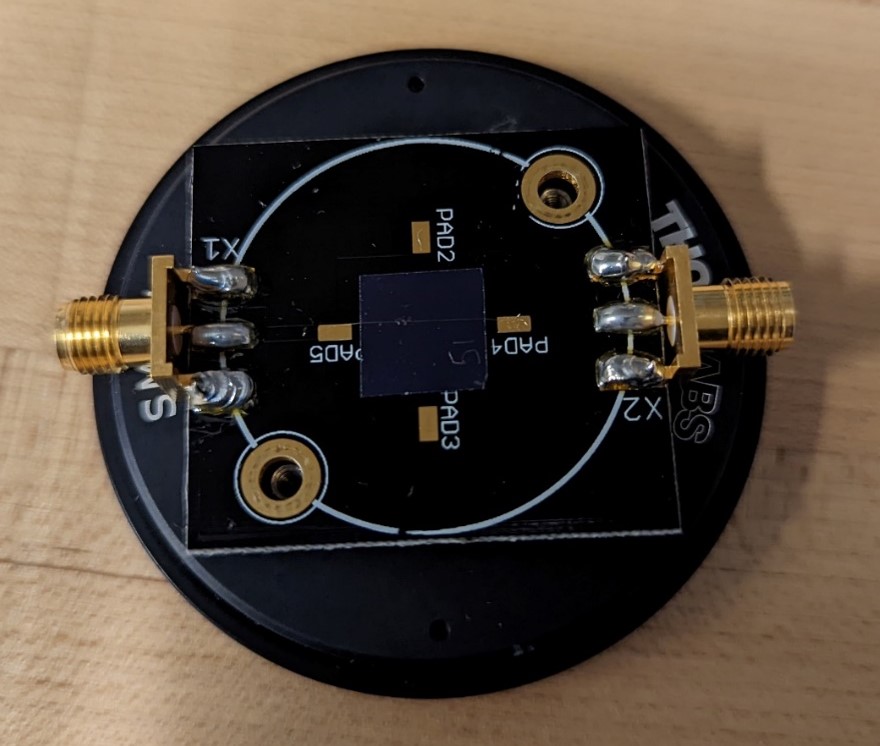}
    \caption{\textbf{Microwave chip}
    }
    \label{fig:microwavechip}
\end{figure}

\section{Measurements} 

\subsection{Photoluminescence characterization}

We raster a fast steering mirror to acquire \textmu-PL images of the h-BN flake and isolated SPEs by recording the counts at each pixel.
The signal and background of a single-photon emitter (SPE) is determined from a two-dimensional Gaussian fit to its \textmu-PL image.
For PL saturation curves, the steady-state PL signal is acquired as a function of excitation power and fit using an empirical saturation model,
\begin{equation}
    \renewcommand{\theequation}{S\arabic{equation}}
    \setcounter{equation}{1}
    C(P) = \frac{C_{\mathrm{sat}}P}{P + P_{\mathrm{sat}}}
    \label{eq:saturationCurve}
\end{equation}
where 
$C$ is the background-subtracted, steady-state PL count rate,
$P$ is the optical excitation power,
$C_{\mathrm{sat}}$ is the saturation count rate, 
and $P_{\mathrm{sat}}$ is the corresponding saturation power.
The polarization scans are acquired to measure the linear excitation and emission polarization properties.
The measurements are acquired by varying the linear polarization of the excitation laser or by passing the PL through a linear polarizer placed in the collection path.
The polarization dependent PL signal is determined by recording the steady-state PL of the SPE at each polarization angle and subtracting the background PL measured at a spatial location offset $\sim$1 \textmu m from the SPE.
A randomized order of the polarization angles minimizes effects of drift and hysteresis.
For excitation polarization measurements, the linear polarizer in the collection path is removed.
For emission polarization measurement, the excitation polarization is set to maximize the PL.
The data are fit using the model function
\begin{equation}
    \renewcommand{\theequation}{S\arabic{equation}}
    I_{s}(\theta) = A_{s} \cos^{2}(\theta-\theta_{s}) + B_{s} \label{eqn:dipole}
\end{equation}
where $s$ indicates excitation (ex) or emission (em), $A_{s}$ is the amplitude, $\theta_{s}$ is the polarization angle of maximum intensity, and $B_{s}$ is the offset.
From the fit results, the visibility is calculated as 
\begin{equation}
    \renewcommand{\theequation}{S\arabic{equation}}
    V_{s} = \frac{I^\mathrm{max}_s-I^\mathrm{min}_s}{I^\mathrm{max}_s+I^\mathrm{min}_s} = \frac{A_{s}}{A_{s} + 2B_{s}} \label{eqn:visibility}    
\end{equation}
where $I^{\mathrm{max}}_s$ and $I^{\mathrm{min}}_s$ are the maximum and minimum PL signal, respectively.
The misalignment between the excitation and emission polarization angles is
\begin{equation}
    \renewcommand{\theequation}{S\arabic{equation}}
    \Delta\theta = \theta_{\mathrm{ex}} - \theta_{\mathrm{em}}
    \label{eq:polarization_alignment}
\end{equation}
The PL spectra are collected as multiple exposures and averaged after correcting for dark counts, cosmic rays and wavelength-dependent photon collection efficiency.
The PL spectra are measured as a function of wavelength, $\lambda$ and binned to determine spectral distribution function, $S(\lambda)$.

\begin{figure}[ht]
    \renewcommand{\thefigure}{S\arabic{figure}}
    \renewcommand\figurename{Figure }
    \let\nobreakspace\relax
    \centering
    \includegraphics[width=2in]{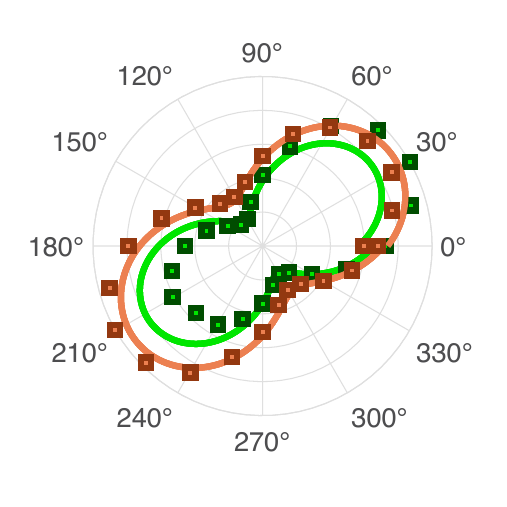}
    \caption{
    \textbf{Emission Polarization}  
    Polar plot showing polarization dependence of emission for 532 nm (shown in green) and 592 nm (shown in orange) excitation.
    }    \label{fig:EmissionPolarization}
\end{figure}
PL intensity as a function of excitation polarization angle is shown in Fig.~\ref{fig:summary}, and polarization-dependent emission is shown in Fig~\ref{fig:EmissionPolarization}. The excitation and emission dipoles are aligned for both green (532 nm) and orange (592 nm) excitation.

\subsection{Photon emission correlation spectroscopy (PECS)}

All PECS, ODMR, time-domain, and spin dynamics measurements were performed under 592 nm illumination.
We calculate $\autocorr{\tau}$ from the photon arrival times acquired from two detectors in a Hanbury Brown and Twiss interferometer using a time-correlated single-photon counting module.
We fit the background-corrected data using a general empirical model for a QE's optical dynamics with a varying number of levels:
\begin{equation}
    \renewcommand{\theequation}{S\arabic{equation}}
    \autocorr{\tau} = 1 - C_{1}e^{-\gamma_{1}|\tau|} + \sum_{i=2}^{n}C_{i}e^{-\gamma_{i}|\tau|}
    \label{eq:generalizedG2}
\end{equation}
Here, $\gamma_{1}$ is the antibunching rate, $C_1$ is the antibunching amplitude, $\gamma_{i}$ for $i\ge 2$ are bunching rates, and $C_i$ for $i\ge 2$ are the corresponding bunching amplitudes.
We determine the number of resolvable timescales, $n$, by calculating and comparing the Akaike information criterion and the reduced chi-squared statistic for each best-fit model, as discussed in Ref. \cite{Patel2022} including background correction procedure.

For a given emitter, all autocorrelation measurements are performed with the excitation polarization set at the angle of maximum excitation and the collection path has the polarizer removed.
Due to the varying emitter brightness, which affects the signal-to-noise ratio of the antibunching signal, measurements are integrated for 10 s to 140 min with repositioning occurring every 2 min.
All errors from the fitting denote 68\% confidence intervals.

Due to timing jitter in the single photon counting modules (SPCMs) introducing systematic artifacts at short delay times, the data are analyzed in two stages: logarithmic and linear scales.
First, the the autocorrelation data are binned over a log scale for visualizing the dynamics over 9 orders of magnitude in time (0.1 ns to 1 s) corrected for background \cite{Brouri2000}, and then fit by multiple instances of Eq.~S\ref{eq:generalizedG2} with $n=[2, 5]$.
The best fit, and corresponding $n$, is then determined by calculating the AIC and comparing the reduced chi-squared statistic.
This method determines the number of bunching levels and their rates and amplitudes that best explain the observations.
The emitter's autocorrelation data are then binned over a linear scale that contains the antibunching features ($\tau \le \SI{30}{\nano\second}$). 
To account for the timing jitter in the SPCMs, the instrument response function (IRF) is found by measuring the autocorrelation signal of an attenuated picosecond pulsed laser sent through the HBT interferometer and binned over the same linear scale as the emitter.
A convolution of the IRF with a modified Eq.~S\ref{eq:generalizedG2} is fit to the background-corrected data, given by
\begin{equation}
    \renewcommand{\theequation}{S\arabic{equation}}
    \tilde{g}^{(2)}(\tau) = \mathrm{IRF} * (1 - C_1 e^{-\gamma_1|\tau|} + C_\mathrm{B}(\tau))
    \label{eq:linear_g2}
\end{equation}
where $C_\mathrm{B}(\tau)$ is the total bunching contribution found from the log scale analysis (first step) and only $C_1$ and $\gamma_1$ are allowed to vary.
The autocorrelation at zero delay is then given by
\begin{equation}
    \renewcommand{\theequation}{S\arabic{equation}}
    \tilde{g}^{(2)}(0) = 1 - C_1 + \sum_{i=2}^n C_i
    \label{eq:g2_0}
\end{equation}
which is used to determine the purity of single-photon emission from the emitter.

\subsection{Optical power and magnetic-field-dependent PECS}
\label{sec:field-dependent-PECS}

\begin{figure}[ht!]
    \renewcommand{\thefigure}{S\arabic{figure}}
    \renewcommand\figurename{Figure }
    \let\nobreakspace\relax
    \centering
    \includegraphics[width=3.375in]{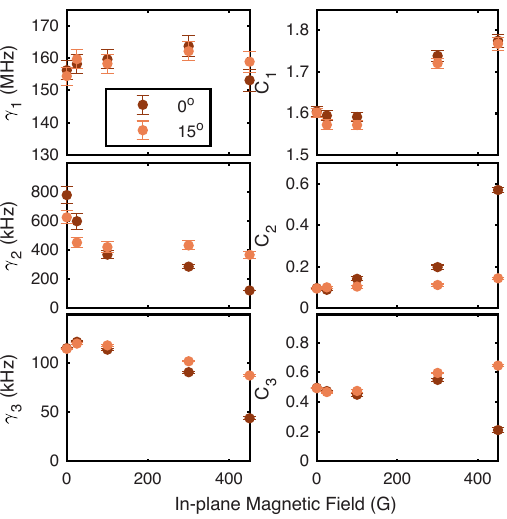}
    \caption{
    \textbf{Magnetic-Field-Dependent Optical Dynamics.}  
    Result of PECS measurements as a function of an in-plane magnetic field for 0$\degree$ and 15$\degree$ dipole orientation.
    All error bars represent 68\% confidence intervals.
    }
    \label{fig:opticalDynamics}
\end{figure}

The magnetic-field-dependent optical dynamics can be probed via observed changes in the temporal dynamics of photon correlations ($\autocorr{\tau}$) as a function of applied magnetic fields.
These changes manifest through variation in the bunching rates and associated amplitudes that arise from inter-system crossings involving spin-selective transitions.
Figure~\ref{fig:opticalDynamics} shows optical dynamics as a function of in-plane magnetic field strength.
We determine the temporal dynamics by fitting $\autocorr{\tau}$ using the empirical function in Eq.~\autocorrEq~at various magnetic field amplitudes.
We apply a magnetic field parallel to the sample and thus to the h-BN with an assumption of it lying flat on the substrate.
We perform the measurement at various dipole orientations.
The antibunching rate, $\gamma_{1}$ stays similar on varying the magnetic field strength, with the average $\gamma_{1}=158$ MHz for both orientations (Fig.~\ref{fig:opticalDynamics}).
The bunching rates, $\gamma_{2}$ and $\gamma_{3}$ vary by more than 5$\times$ with the magnetic field strength, albeit to different extents for different dipole orientations (Fig.~\ref{fig:opticalDynamics}).
The total bunching amplitude, $C_{2}+C_{3}$ increases by over 30\% on increasing the magnetic field strength from 0 G to 470 G.
These observations confirm the magnetic-field-dependent transitions indicating presence of spin in the SPE at room temperature.

\begin{figure}[htb!]
    \renewcommand{\thefigure}{S\arabic{figure}}
    \renewcommand\figurename{Figure }
    \let\nobreakspace\relax
    \centering
    \includegraphics[width=3.375in]{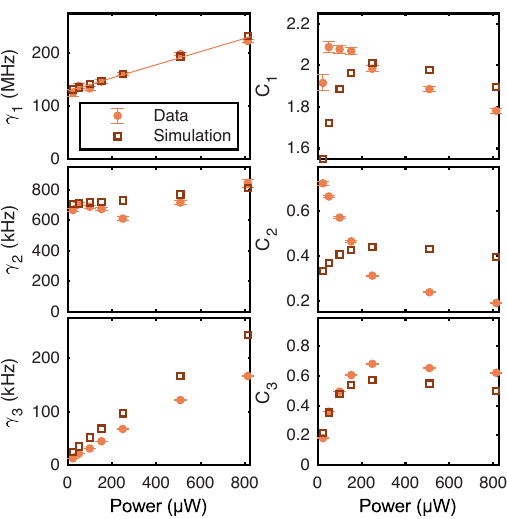}
    \caption{
    \textbf{Excitation-Power-Dependent Optical Dynamics with Simulation Results.}  
    Full result of PECS measurements as a function of excitation power for 0$\degree$ dipole orientation and 0 G magnetic field.
    Bunching amplitudes and simulation results (right column) are shown here alongside the rates shown in the main text.
    Error bars represent 68\% confidence intervals.
    Orange line shows linear fit of $\gamma_{1}$ data as a function of optical excitation power.
    The fit gives a y-intercept of $128 \pm 2$ MHz and a slope of $0.125 \pm .005$ MHz/\textmu W.
    }
    \label{fig:simResults}
\end{figure}

The full excitation-power-dependent PECS are presented in Fig.~\ref{fig:simResults} with comparison to simulated dynamics using the transition rates discussed in the main text.
Overall power-dependent trends and orders of magnitude are consistent between the simulation and data for all rates and for bunching amplitudes $C_1$ and $C_3$.
The slight variations are due to the parameter optimization at 91 G magnetic field for the fluorescence recovery experiment, while the PECS power-dependent data was taken at 0 G field.
Applied magnetic field can impact spin-dependent transition rates.
The discrepancy between the data and simulation for bunching amplitude $C_2$ suggests that additional model complexity may be required to capture the full dynamics of the system.

\subsection{Optically detected magnetic resonance}

\begin{figure*}[ht!]
    \renewcommand{\thefigure}{S\arabic{figure}}
    \renewcommand\figurename{Figure }
    \let\nobreakspace\relax
    \centering
    \includegraphics[width=6.6in]{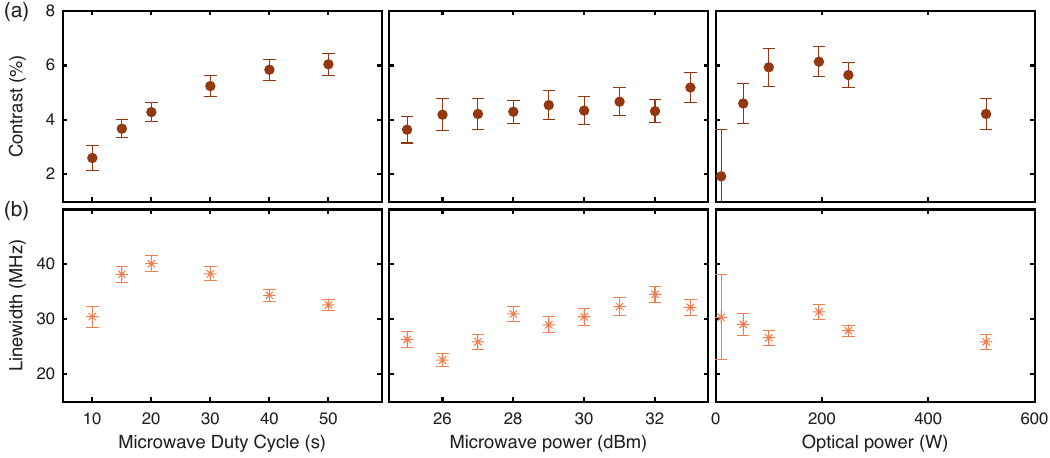}
    \caption{
    \textbf{Optically detected magnetic resonance contrast and linewidth.}      
    \textbf{(a)} The cw ODMR contrast and 
    \textbf{(b)} the linewidth as a function of microwave pulse duration, microwave power and optical power for 470 G in-plane magnetic field and 0$\degree$ dipole orientation. 
    All error bars represent 68\% confidence intervals.
    }
    \label{fig:odmr_optimized}
\end{figure*}

\begin{figure}[hb!]
    \renewcommand{\thefigure}{S\arabic{figure}}
    \renewcommand\figurename{Figure }
    \let\nobreakspace\relax
    \centering
    \includegraphics[width=3.375in]{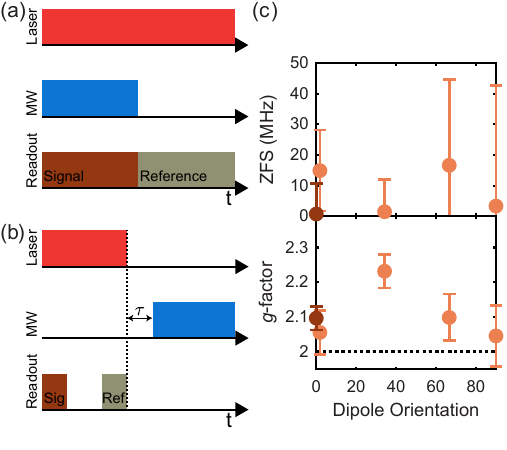}
    \caption{
    \textbf{Optically detected magnetic resonance.}      
    \textbf{(a)} Pulse protocol for cw ODMR.
    \textbf{(b)} Pulse protocol for pulsed ODMR. 
    \textbf{(c)} The zero-field splitting (ZFS) and $g$-factor as a function of dipole orientation.
    The light (dark) orange data are obtained from pulsed (cw) ODMR.
    All error bars represent 68\% confidence intervals.
    }
    \label{fig:ZFS}
\end{figure}

Cw ODMR data in the main text was taken using the pulse protocol shown in Fig.~\ref{fig:ZFS}(a) with a 40 \textmu s microwave pulse followed by a 40 \textmu s wait during continuous 592 nm optical excitation at 200 \textmu W (Fig.~\ref{fig:summary}(e)) and 500 \textmu W (Fig.~\ref{fig:Spin}(a-b)) optical power.
Figure~\ref{fig:odmr_optimized} presents cw ODMR contrast and linewidth as a function of microwave pulse duration, microwave power, and optical power.
Each cw ODMR curve is fit to a Lorentzian function with a y-offset fixed to 1, and the peak contrast and linewidth are extracted from the fit.
The Lorentzian function used is
\begin{equation}
    \renewcommand{\theequation}{S\arabic{equation}}
    L(f) = 1 + \frac{B}{(f-f_{o})^{2} + \Gamma^{2}}
    \label{eq:lorentzian}
\end{equation}
where $f$ is the microwave frequency, $B$ is a constant, $f_{o}$ is the central or resonance frequency, and $\Gamma$ is the half-width at half-maximum (HWHM) or linewidth.
The fit is weighted by errors propagated from Poisson noise.
Percent contrast at each frequency is calculated as
\begin{equation}
C(f) = 100 \times (L(f) - 1).
\end{equation}

At an in-plane applied magnetic field, we sweep amplified microwave frequencies using a signal generator.
An arbitrary waveform generator controls optical and microwave pulses. 
The microwave pulse duration has a strong effect on the contrast (Fig.~\ref{fig:odmr_optimized}(a)), increasing it by 3$\times$ on increasing the pulse duration from 10 \textmu s to 50 \textmu s, keeping the linewidth (Fig.~\ref{fig:odmr_optimized}(b)) $\sim$30 MHz.
Below 10 \textmu s pulse duration, the contrast is almost zero. 
With optical power, we see a non-monotonic change in contrast and linewidth and find the optimum power to be in the range 150 \textmu W to 300 \textmu W for maximum contrast.

\begin{figure}[hb!]
    \renewcommand{\thefigure}{S\arabic{figure}}
    \renewcommand\figurename{Figure }
    \let\nobreakspace\relax
    \centering
    \includegraphics[width=3.375in]{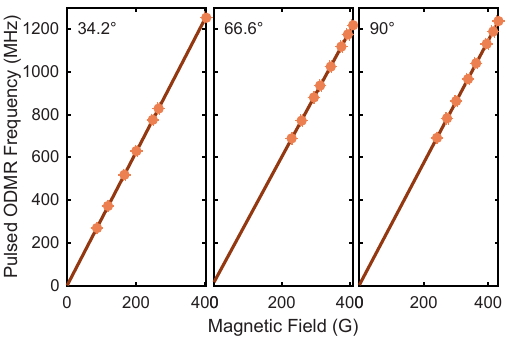}
    \caption{
    \textbf{Resonance frequency at various dipole orientations.} 
    The circles represent resonance frequency for a given dipole orientation at an in-plane magnetic field, determined using the Lorentzian function (Eq. ~S\ref{eq:lorentzian}) discussed in the text.
    The data were acquired at 350 \textmu W optical power and 28 dBm microwave power for 34.2$\degree$ orientation and 1 W for 66$\degree$ and 90$\degree$ orientation.
    The solid line is a fit to the data using a linear function (Eq. ~S\ref{eq:linearFit}).
    The error bars are same as the size of the data points and represent 68\% confidence intervals.
    }
    \label{fig:dipoleDependentZFS}
\end{figure}

Upon optimizing the parameters of cw ODMR, we use pulsed ODMR to probe ZFS, $g$-factor and hyperfine interactions (Fig.~\ref{fig:summary}(f)).
Figure~\ref{fig:ZFS}(b) presents the pulse protocol for pulsed ODMR measurement.
A 40 \textmu s laser pulse is followed by a 20 \textmu s wait and 40 \textmu s microwave pulse.
During the laser pulse, the readout consists of signal during first 5 \textmu s and reference during last 5 \textmu s.
The pulse durations and optical power used were obtained from measurements to maximize the SNR.
The pulsed ODMR contrast is obtained by normalizing signal with reference.
The ZFS and $g$-factor obtained from the fits are presented in Fig.~\ref{fig:ZFS}(c).
Figure~\ref{fig:dipoleDependentZFS} presents resonance frequency as a function of in-plane magnetic field for different dipole orientations, extension of data presented in Fig.~\ref{fig:ZFS}(c).
A linear function accounting for uncertainty in frequency and magnetic field is fit to the data,
\begin{equation}
    \renewcommand{\theequation}{S\arabic{equation}}
    R(M) = R_{0} + s \times M
    \label{eq:linearFit}
\end{equation}
where $M$ is the magnetic field, $s$ is the slope and $R_{0}$ is the y-intercept.
The y-intercept corresponds to the ZFS, and $s$ divided by Bohr magneton gives the $g$-factor.

\subsection{Signal-to-noise ratio}


We record optical spin contrast using the pulse protocol shown in Fig.~\ref{fig:odmr}(b) 
The delay between signal and reference readout, $\tau_\mathrm{D}$, is the same as $\tau_\mathrm{w}$ plus microwave pulse duration.
For the data in Fig.~\ref{fig:odmr}(c), we set the initialization optical pulse to be 40 \textmu s at 350 \textmu W, setting wait time $\tau_\mathrm{w}$ to be 20 \textmu s and microwave pulse length to be 40 \textmu s.
Figure~\ref{fig:odmr}(c) shows a representative optical spin contrast obtained from measurement using this pulse protocol.
The green highlighted region corresponds to an example of a readout window, $\tau_\mathrm{R}$, from which $\alpha_1$ and $\alpha_0$ are determined.

\begin{figure*}[]
    \renewcommand{\thefigure}{S\arabic{figure}}
    \renewcommand\figurename{Figure }
    \let\nobreakspace\relax
    \centering
    \includegraphics[width=7in]{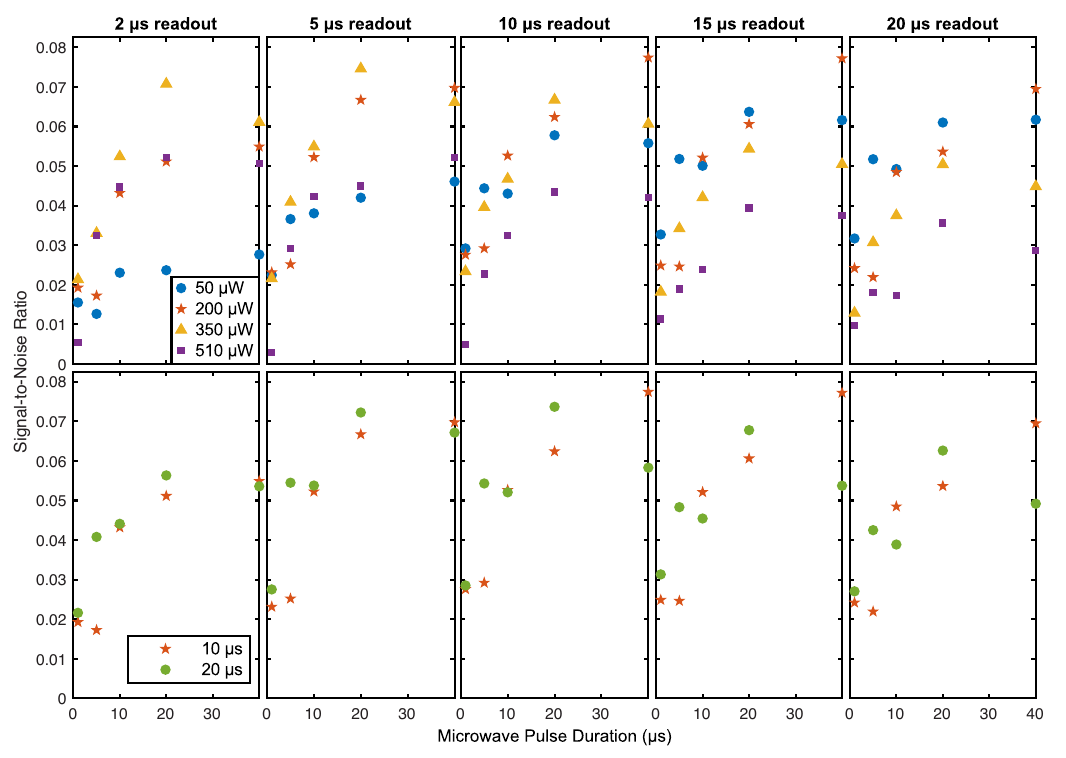}
    \caption{
    \textbf{Signal-to-Noise Ratio.}      
    SNR determined from optical spin contrast measurements as a function of microwave pulse duration for different readout times for \textbf{(top row)} varying optical power (with $\tau_\mathrm{w} = 10$ \textmu s fixed) and \textbf{(bottom row)} varying wait time (with $p = 200$ \textmu W fixed).
    Data was acquired at 592 nm optical excitation and 34 dBm microwave power.
    }
    \label{fig:snr}
\end{figure*}

We record optical spin contrast curves, varying the optical power (Fig.~\ref{fig:Spin}(c)), the wait time, and the microwave pulse duration. 
From each optical spin contrast curve, we calculate SNR for $\tau_\mathrm{R}$ in the range of 2 \textmu s to 20 \textmu s.
Figure~\ref{fig:snr} presents the SNR for various readout times and all the optical powers, microwave pulse durations and wait times probed.

\section{Simulations}

\subsection{Energy-level model for simulations}
\begin{figure}[htb!]
    \renewcommand{\thefigure}{S\arabic{figure}}
    \renewcommand\figurename{Figure }
    \let\nobreakspace\relax
    \centering
    \includegraphics[width=3.375in]{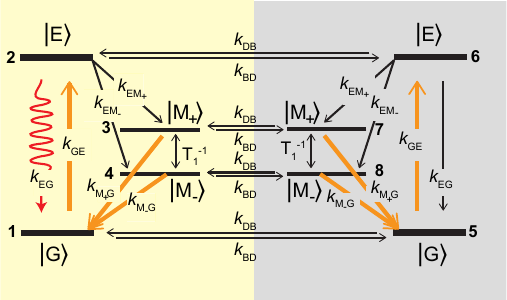}
    \caption{
    \textbf{Optical Dynamics Model for Simulation.}  
     Eight-energy-level model used in simulations to capture modulation of the emission mechanism between a radiative (yellow box) and non-radiative (grey box) configuration.
    Each configuration contains a singlet ground and excited state, and a metastable doublet.
    Black arrows depict non-radiative power-independent transitions, orange arrows depict excitation-power-dependent transitions, and the red wiggly arrow depicts the radiative transition.
    }
    \label{fig:8lvl}
\end{figure}
Our model shown in Fig.~\ref{fig:odmr}(a) contains four energy levels and an emission modulation mechanism. In order to simulate this model, we index the energy levels by $i$ = 1, 2, 3, or 4, and we index the emission mechanism by $j$ = 1 or 2.
We represent these eight unique state combinations using an eight-level model shown in Fig.~\ref{fig:8lvl} with each state labeled 1-8.
The model features two manifolds, one with a radiative emission mechanism ($j=1$) and the other with a non-radiative emission mechanism ($j=2$).
The emission modulation dynamics are captured in the simulation by the ability to switch between the two manifolds at any time at rates $k_\mathrm{DB}$ and $k_\mathrm{BD}$.
Each manifold has identical electronic states ($i = 1-4$) and optical dynamics to ensure that the state evolution is independent of the emission modulation.
Time-dependent PL is assumed to be proportional to the time-dependent excited-state population in the radiative manifold.
Therefore, the fluorescence recovery simulation is normalized by the simulated steady-state excited state population in the j=1 manifold and by the experimentally observed steady-state PL.

\subsection{Optical dynamics simulations}
In order to simulate the classical evolution of state populations for the model shown in Fig.~\ref{fig:odmr}(a), we use the rate equation,
\begin{equation}
    \dot{P} = GP,
    \label{eq:master}
\end{equation}
where $P$ is a vector of state occupation probabilities and $G$ is a matrix of all transition rates between states.
The results of our simulations are displayed in Figs.~\ref{fig:odmr}(d-f) and \ref{fig:DSR}(c).
With initial conditions set by the experimental protocol, we use MATLAB’s ODE solver, \texttt{ode15s}, to solve Eq.~S\ref{eq:master} for $P(t)$, the time-dependent state probabilities.
For PECS simulations, the initial condition is the state of the system immediately following the release of a photon, $P_{|G\rangle}(0) = 1$.
For the fluorescence recovery and spin contrast experiments, the initial condition is the steady-state, which is calculated by numerically solving the steady-state equation,
\begin{equation}
    0 = GP.
\end{equation}
We simulate optical and microwave pulses by modifying the corresponding affected rates in the transition rate matrix, $G$.
For pulse sequences that involve turning the laser off, we set the optical excitation rate to zero, $k_{GE} = 0$ and we set the pumped metastable decay rate, $k_{MG}$ to be equal to its spontaneous component only.
When the laser is turned on, those rates are returned to their original values.
For protocols that involve turning microwaves on, we set the rate between the two metastable spin states based on the Rabi frequency, and restore that rate to $T_{1}^{-1}$ when microwaves are turned off.
In order to simulate experiments with multiple pulses, we start the population distribution at the experiment's initial condition (specified above), and solve Eq.~S\ref{eq:master} according to the length of the first part of the sequence. We record the interim population distributions at the end of each pulse to be carried over as the initial conditions for the start of the next part of the sequence.
For each simulation, we track the time-dependent excited state population to capture relative changes in time-dependent PL.

PECS simulation code and example can be found at https://github.com/penn-qel/photon-emission-correlation-spectroscopy.
More detail on optical dynamics simulations can be found in Ref.~\cite{Fishman2023}.

\subsection{Quantifying transition rates}

In order to separate the system's overall optical dynamics from its features that determine spin polarization, we parameterize the metastable transition rates as overall rates and spin branching ratios.
The transition rates from the excited to metastable states ($k_\mathrm{EM_\pm}$) are parameterized as the total excited-to-metastable transition rate, $k_\mathrm{EM}\equiv k_\mathrm{EM_{-}}+k_\mathrm{EM_{+}}$, with branching ratio $k_\mathrm{EM_{+}} / k_\mathrm{EM_{-}}$.
The transition rates from the metastable states to the ground state ($k_\mathrm{M_\pm G}$) are parameterized as the average metastable-to-ground transition rate, $k_\mathrm{MG}\equiv(k_{\mathrm{M_{+}G}}+k_{\mathrm{M_{-}G}})/2$, with branching ratio $k_{\mathrm{M_{+}G}} / k_{\mathrm{M_{-}G}}$.

To quantify the transition rates between each electronic state in our model, Fig.~\ref{fig:8lvl}, we simulate a fluorescence recovery experiment (Fig.~\ref{fig:DSR}(a)), while systematically varying transition rates to match features observed in Fig.~\ref{fig:DSR}(c).
The resulting rates are discussed in the main text.
The first feature we consider is the time constant that describes the increase in maximum PL at the start of the laser pulse as a function of $\tau_\mathrm{D}$ (dashed line in Fig.~\ref{fig:DSR}(c)).
This time constant is determined by the branching ratio $k_{\mathrm{M_{+}G}} / k_{\mathrm{M_{-}G}}$ and the $p$-independent component of $k_{\mathrm{MG}}$.
We simulate and fit fluorescence recovery, varying those two parameters to find combinations of $k_{\mathrm{M_{+}G}} / k_{\mathrm{M_{-}G}}$ and $k_{\mathrm{MG}}$ that give the correct time constant.
Next, we consider the maximum contrast from the initial to the steady-state PL, which depends on the ratio, $k_{\mathrm{MG}}/k_{\mathrm{EM}}$.
We simulate fluorescence recovery, varying $k_{\mathrm{MG}}$ and $k_{\mathrm{EM}}$ to find which ratio gives the correct contrast.
Finally, we consider the decay time from maximum to steady-state PL during the laser pulse.
This decay time is dependent on many rates.
However, taking into account the constraints generated by the first two simulation steps, we can simulate this decay to steady-state in the regime that gives the correct maximum contrast and time constant to empirically determine a value for $k_{\mathrm{EM}}$.
We draw upon constraints from PECS data (as described in the main text) and make use of PECS and spin contrast simulations to gain further clarity on any remaining parameters such as fluorescence modulation rates ($k_\mathrm{BD}$ and $k_\mathrm{DB}$) and the branching ratio $k_\mathrm{EM_{+}} / k_\mathrm{EM_{-}}$.

\subsection{Simulations of microwave power-dependent ODMR}

We now provide more detail regarding the ODMR contrast simulations, as a function of microwave power and frequency, shown as solid lines in Fig.~\ref{fig:Spin}(a). We simulate ODMR by solving the Lindblad master equation:
\begin{equation}
\renewcommand{\theequation}{S\arabic{equation}}
\label{eq:master equation}
    \partial_{t}\hat{\rho}(t)=-\frac{i}{\hbar}\left[\hat{H}(t),\hat{\rho}(t)\right]+\sum_{i}\hat{\mathcal{L}}_{i}\left[\hat{\rho}(t)\right]
\end{equation}
with 
\begin{equation}
\renewcommand{\theequation}{S\arabic{equation}}
\label{eq:Lindbladian}
   \hat{\mathcal{L}}_{i}\left[\hat{\rho}(t)\right]\equiv k_{i}\left(\hat{L}_{i}\hat{\rho}(t)\hat{L}_{i}^{\dagger}-\frac{1}{2}\left\{\hat{L}_{i}^{\dagger}\hat{L}_{i},\hat{\rho}(t)\right\}\right)
\end{equation}
where the Hamiltonian, $\hat{H}(t)$, is block-diagonal in the ground, excited, and metastable manifold Hamiltonians. The Lindblad equation tracks the coherent dynamics of the metastable spin states while approximating the transitions between ground, excited, and metastable manifolds as decoherent processes. This treatment allows for a quantitative simulation of the excited manifold population, and its dependence on the dynamics of the metastable spin states. Here, the ground and excited manifolds are treated as singlet states that do not respond to magnetic fields, while the Hamiltonian of the metastable manifold has the form:
\begin{equation}
    \renewcommand{\theequation}{S\arabic{equation}}
    \label{eq:Hamiltonian}
    \hat{H}_{m}=g\mu_{B}\left(\vec{B}_{0}+\vec{B}_{1}(t)\right)\cdot\hat{\vec{S}}
\end{equation}
where $g$ is the isotropic $g$-factor, $\mu_{B}$ is the Bohr magneton, $\vec{B}_{0}$ is the in-plane magnetic field, $\vec{B}_{1}(t)$ is the microwave magnetic field, and $\hat{\vec{S}}$ is the spin of the emitter ($S=\frac{1}{2}$).

The superoperators, $\hat{\mathcal{L}}_{i}\left[\hat{\rho}(t)\right]$, account for hopping processes between manifolds such as absorption, emission, and non-radiative relaxations through the metastable manifold. Spin dephasing and spin relaxation within manifolds is also simulated using Lindblad jump operators, $\hat{L}_{T_{2}^{*}}=\hat{\sigma}_{z}$ and $\hat{L}_{T_{1}}=\hat{\sigma}_{x}$, with the respective coherence times related to the inverse of the rate constants, $k_{T_{2}^{*}}$ and $k_{T_{1}}$. The first term in (S\ref{eq:Lindbladian}) generates populations in the final state, while the second term dissipates population in the initial state. With this insight, we define PL as:
\begin{equation}
\renewcommand{\theequation}{S\arabic{equation}}
\label{eq:PL}
    PL(B_{1},f)\equiv Tr\left[\hat{L}\hat{\rho}(\tau_{ss})\hat{L}^{\dagger}\right]
\end{equation}
and PL contrast:
\begin{equation}
\renewcommand{\theequation}{S\arabic{equation}}
\label{eq:PL contrast}
    \frac{\Delta PL}{PL}\left(f\right)\equiv\frac{PL(B_{1},f)}{PL(0,f)}
\end{equation}
where the operator in (S\ref{eq:PL}) extracts the value of the excited state population in the steady-state and $f$ is the microwave frequency. However, as stated in the main text, nearly all input parameters $k_{i}$ in the Lindblad master equation were determined empirically from the PECS and time-domain data, and we note there may be some amount of uncertainty in these parameters.
To account for these uncertainties, we introduce an amplitude scaling factor into the PL contrast expression:
\begin{equation}
\renewcommand{\theequation}{S\arabic{equation}}
\label{eq:PL contrast with Amps}
    \frac{\Delta PL}{PL}\left(f,A\right)\equiv1+A\left(\frac{\Delta PL}{PL}\left(f\right)-1\right)
\end{equation}
The justification for $A$ comes from the insight that for the optical transition rates and Rabi frequencies of the ODMR measurement, uncertainties in $T_{1}$ have little effect on the FWHM of the ODMR contrasts. Therefore, by setting $T_{1}$ = 100 \textmu s and introducing the fitting parameter $A$, we minimize the amplitude discrepancies between the ODMR data and simulations. This allows for an accurate determination of $T_{2}^{*}$ and $\Omega_{R}$, which predominately affect the FWHM, through fitting. In this way, through a simultaneous fit of $T_{2}^{*}$ and the microwave power to microwave magnetic field coupling parameter $\eta$, where $\Omega_\mathrm{R}/(2\pi)=\eta \sqrt{p_\mathrm{MW}}$, and individual fits of $A$ and $g$ for each microwave power data set, we were able find a set of best-fit parameters with a reduced chi-squared $\chi^{2}_{r}$ = 1.94. The results of ODMR simulations with these best-fit parameters for $T_{1}$ = 100 \textmu s are depicted in Fig.~\ref{fig:Spin}(a). We also repeated this fitting process for $T_{1}$ = 50 and 200 \textmu s to quantitatively probe the dependence of the ODMR contrast FWHM on $T_{1}$, and the results are shown in Table~\ref{tab:fitting params}. We see that the best-fit $T_{2}^{*}$ and $\eta$ are virtually unchanged when the amplitude scaling factors are included, affirming our insight that $T_{1}$ has little effect on the FWHM of the ODMR contrast. We conclude that the values of $T_{2}^{*}$ and $\eta$, which we obtained through fitting the microwave power-dependent ODMR data to our model (Eqs.~S\ref{eq:master equation}-S\ref{eq:PL contrast with Amps}), are tightly constrained.
We calculate parameter uncertainties from both the fit confidence intervals and the variation in best-fit values due to possible variation in $T_1$ (see Table~\ref{tab:fitting params}).
The uncertainties quoted in the main text result from the quadrature addition of uncertainties from these two sources.
\newline
\begin{table}[h]
        \begin{tabular}{|c|c|c|c|}
        \hline
        \rule{0pt}{3ex}
         $T_{1}$ (\textmu s) & $T_{2}^{*}$ (ns) & $\eta$ (MHz/$\sqrt{\mathrm{W}}$) & $\chi^{2}_{r}$ \\[1ex] 
         \hline
         \rule{0pt}{3ex}
         200 & 6.2 $\pm$ 0.1 & 0.0184 $\pm$ 0.0001 & 1.95 \\[1ex] 
         \hline
         \rule{0pt}{3ex}
         100 & 6.3 $\pm$ 0.1 & 0.0189 $\pm$ 0.0001 & 1.94 \\[1ex]
         \hline
         \rule{0pt}{3ex}
         50 & 6.3 $\pm$ 0.1 & 0.0199 $\pm$ 0.0001 & 1.94\\[1ex]
         \hline
        \end{tabular}
    \caption{Best-fit $T_{2}^{*}$ and $\eta$ for Different Choices of $T_{1}$.}
    \label{tab:fitting params}
\end{table}

\section{Additional discussion}
\subsection{Location of metastable doublet}

In the main text, we briefly discuss how our observations in the spin contrast experiment (Fig.~\ref{fig:odmr}(c)) support a metastable doublet.
Here, we provide further intuition for this conclusion.
First we note the positive ODMR contrast observed (Fig.~\ref{fig:summary}(e)).
For the configuration with spin doublets in the optical manifold, positive ODMR contrast implies that the laser polarizes the spin to favor the spin state with higher probability of decay into the metastable state. 
A resonant microwave field depolarizes the spin, decreasing the probability of decaying through the metastable state, increasing the PL intensity.
In a spin contrast measurement, this lower non-radiative decay probability manifests as a slower decay in the signal PL from its peak to its steady state, compared to that of the reference.
Meanwhile, the microwave does not affect the metastable state's decay rate to the optical ground state, so the initial PL intensity remains unchanged.
We would therefore expect to see the signal and reference signals in Fig.~\ref{fig:odmr}(c) begin together and then diverge at longer delay times.

On the other hand, for the metastable spin doublet configuration, positive ODMR contrast implies that optical excitation polarizes the spin to favor the slower-decaying metastable spin state.
The microwave pulse increases the probability of decay from the metastable state to the ground state.
We therefore expect to observe a higher initial PL for the signal compared to the reference, followed by decay to the steady state at identical rates.
Our results in Fig.~\ref{fig:odmr}(c) show significant spin contrast within the first 10 \textmu s of readout, followed by a similar decay time for both the signal and reference, consistent with the metastable doublet configuration.




\end{document}